\documentclass{iaus}
\usepackage{graphicx}

\title[Transit Detection of RV Planets] 
{Transit Detection of Radial Velocity Planets}

\author[S.R. Kane \& K. von Braun]   
{Stephen R. Kane \and Kaspar von Braun}

\affiliation{Michelson Science Center, Caltech, MS 100-22,
770 South Wilson Avenue Pasadena, CA  91125, USA\\
email: {\tt skane@ipac.caltech.edu}}

\pubyear{2008}
\volume{253}  
\pagerange{1--4}
\jname{Transiting Exoplanets}
\editors{Editor, eds.}
\begin{document}

\maketitle


\begin{abstract}

The orbital parameters of extra-solar planets have a significant
impact on the probability that the planet will transit the host
star. This was recently demonstrated by the transit detection of HD
17156b whose favourable eccentricity and argument of periastron
dramatically increased its transit likelihood. We present a study
which provides a quantitative analysis of how these two orbital
parameters effect the geometric transit probability as a function of
period. Further, we apply these results to known radial velocity
planets and show that there are unexpectedly high transit
probabilities for planets at relatively long periods. For a
photometric monitoring campaign which aims to determine if the planet
indeed transits, we calculate the significance of a null result
and the subsequent constraints that may be applied to orbital
parameters.

\end{abstract}


\firstsection 


\section{Transit Probability}

There have been at least five cases in which planetary transits were
detected through photometric follow-up of planets already known via
their radial velocity (RV) discoveries. The case of HD 17156b
(\cite{bar07a}) is of particular interest since it is a 21.2 day
period planet which happens to have a large eccentricity ($e = 0.67$)
and an argument of periastron which places the periapsis of its orbit
in the direction toward the observer and close to parallel to the line
of sight, resulting in an increased transit probability.

Recent work by \cite{bar07b} and \cite{bur08} showed that higher
eccentricities of planetary orbits will increase their transit
probabilities and, consequently, expected yield for transit
surveys. We demonstrate the combined effect of the eccentricity, $e$,
and argument of periastron, $\omega$, on transit probability. As shown
by \cite{kan07a}, the place in a planetary orbit where it is possible
for a transit to occur (where the planet passes the star-observer
plane perpendicular to the planetary orbit) is when $\omega + f = \pi
/ 2$. The probability of such a transit occurring, $P_t$, is given by
\begin{equation}
  P_t = \frac{(R_p + R_\star)(1 + e \cos (\pi/2 - \omega))}{a (1 - e^2)},
  \label{transit_prob}
\end{equation}
where $R_p$ and $R_\star$ are the radii of the planet and star
respectively, and $a$ is the semi-major axis.  The orbital
configuration, especially with regards to the values of $e$ and
$\omega$, plays a major role in determining the likelihood of a planet
transiting the parent star.


\section{Argument of Periastron Dependence}

As we rotate the semi-major axis of the orbit around the star we can
observe how the transit probability varies using Equation
\ref{transit_prob}. This dependence is shown in Figure 1 for
eccentricities of 0.3 (dashed line) and 0.6 (dotted line) in
comparison with the constant transit probability for a circular orbit
(solid line). Since the shape of this variation is independent of
period, $P$, the y-axes are scaled for both a 4.0 day and 50.0 day
period orbits. Note that $P_t$ scales linearly with the sum of $R_p$
and $R_\star$.

The peak transit probability occurs at $\omega = \pi / 2$, and the
corresponding increase in $P_t$ as compared to a circular orbit can be
significant; a factor of 1.5 for $e = 0.3$ and a factor of 2.5 for $e
= 0.6$. Moreover, the fraction of the orbital path which produces a
higher value of $P_t$ than the circular orbit with the same period
(corresponding to the fraction of range in $\omega$ for which the
dotted or dashed line is above the solid line in Figure 1) increases
with increasing eccentricity.

\begin{figure}
  \begin{center}
    \includegraphics[angle=270,width=9.0cm]{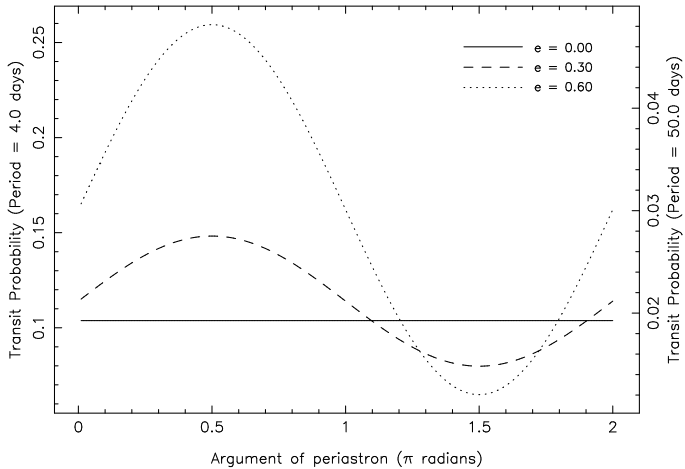}
  \end{center}
  \caption{Dependence of geometric transit probability on the argument
    of periastron, $\omega$, for eccentricities of 0.0, 0.3, and 0.6,
    plotted for periods of 4.0 days (left ordinate) and 50.0 days
    (right ordinate). Stellar and planetary radii are assumed to be a
    Jupiter and solar radius, respectively.}
\end{figure}


\section{Period Dependence}

The current distribution of eccentricities for the known extra-solar
planets indicates that orbits within 0.1~AU tend to be forced into
nearly circular orbits through tidal circularization, whereas longer
period orbits can possess a great range of eccentricities
(\cite{for08}). Indeed most of the planets beyond 0.1~AU have
eccentricities in excess of 0.3.

\begin{figure}
  \begin{center}
    \includegraphics[angle=270,width=9.0cm]{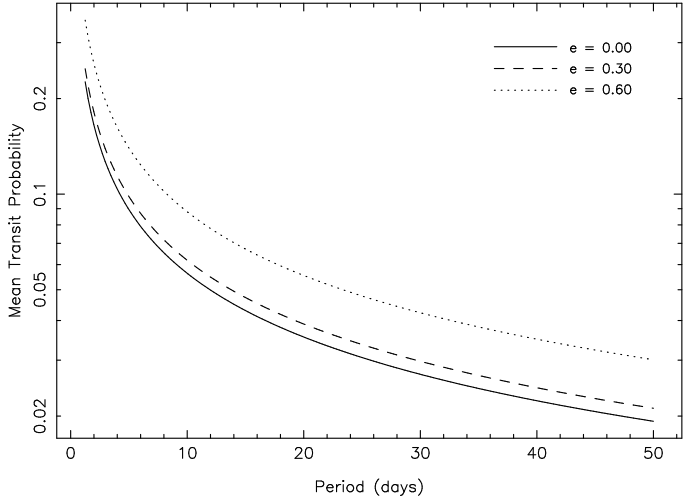}
  \end{center}
  \caption{The mean transit probability on a logarithmic scale,
    averaged over all values of $\omega$, as a function of period, for
    eccentricities of 0.0, 0.3, and 0.6.}
\end{figure}

In Figure 2 we show mean transit probability as a function of period
after averaging over $0 \leq \omega \leq 2 \pi$, for the period range
$1 \leq P \leq 50$ days. Eccentricities of 0.0, 0.3, and 0.6 are shown
with solid, dashed, and dotted lines, respectively. As expected, we
see that doubling the eccentricity from 0.3 to 0.6 creates a
significant increase in the mean transit probability. Most affected
are the longer period planets whose eccentric orbits can raise their
likelihood of transit from a negligible value to a statistically
viable number for photometric follow-up.


\section{Application to Known Exoplanets}

Figure 3 shows the transit probability calculated from orbital
parameters provided by \cite{but06} for planets with estimates of $e$
and $\omega$ (203 planets in total). For the purposes of comparison,
we assume a Jupiter and Solar radius for the values of $R_p$ and
$R_\star$, respectively, and include a solid line which indicates the
transit probability for a circular orbit. In addition, the sub-panel
in the plot shows the difference in $P_t$ between the actual orbit and
a hypothetical circular one of the same period (residuals). The mean
value of the residuals for all 203 planets is positive but relatively
small ($4.13 \times 10^{-5}$), and is dominated by the low transit
probability of the long period planets. The mean residual of planets
with $P < 100.0$ days yields an overall increase of $\sim 0.5$\% in
$P_t$.

\begin{figure}
  \begin{center}
    \includegraphics[angle=270,width=9.0cm]{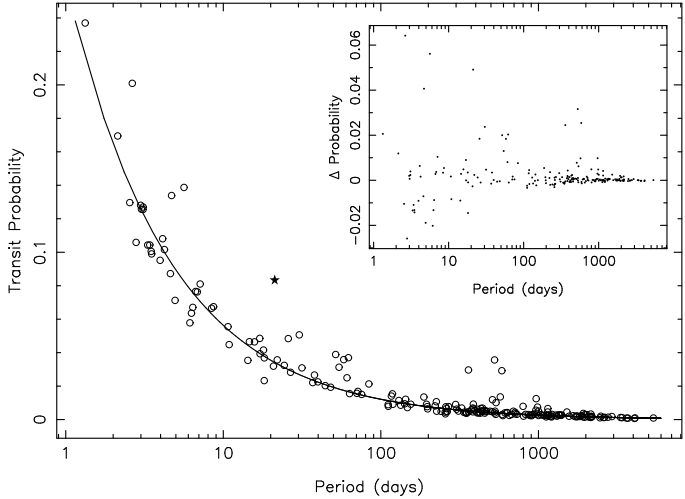}
  \end{center}
  \caption{The geometric transit probability for a circular orbit
    (solid curve) along with the transit probability for 203 RV
    planets from \cite{but06} calculated from their orbital
    parameters (open circles). HD 17156b is indicated by a 5-pointed
    star. The sub-panel plots the difference in $P_t$ between the the
    actual orbit and a hypothetical circular orbit for each of the
    planets.}
\end{figure}

HD~17156b, a transiting planet with 21.2 day period, is shown as a
5-pointed star. Its transit probability is greatly increased by its
orbital parameters. Note that the actual $P_t$ of HD~17156b is larger
than the 5\% shown in Figure 3 since the radius of the host star is
1.47 solar radii. At longer periods, the planets with the largest
residuals are HD~156846b, HD~4113b, and HD~20782b, which have periods
of 359.51, 526.62, and 585.86 days, respectively. The probability
residuals for these three planets are 0.024, 0.032, and 0.025
respectively, the effect of which is to raise their transit
probabilities to the same level as HD~17156b if it were in a circular
orbit.


\section{Global Statistics and Photometric Follow-up}

The majority of radial velocity planets have been detected around $V <
14$ stars. \cite{kan07a} showed that 1.0m class telescopes are ideal
instruments to photometrically monitor these targets. By selecting
observable targets with well constrained transit windows and high
transit probabilities, an optimised campaign can be constructed.

We can determine the significance of a hypothetical null result from a
photometric follow-up campaign by performing a Monte-Carlo simulation
of the transit probabilities calculated from Equation
\ref{transit_prob} and the tabulated planetary and stellar paramaters
of \cite{but06}. Performing this calculation $\sim 10000$ times for
each star produces a probability distribution for the number of
transiting planets expected from this sample, as shown in Figure 4.
The mean value of $\sim 4.5$ transits peaks at $P_t \sim 0.2$ with a
standard deviation of $\sim 2.0$. The probability that none of the
planets in this sample transit their host star is $\sim 1$\%. In fact,
three of the planets in this sample are known to transit, specifically
HD 17156b, GJ 436b, and HD 147506b. Hence the current number of
transiting planets from this sample is almost 1$\sigma$ below the
expectation. This demonstrates that the offset is quantifiable, and
that further transit discoveries in this sample are possible or even
likely which would lead to further understanding of the respective
observational biases of the RV and transit methods.

\begin{figure}
  \begin{center}
    \includegraphics[angle=270,width=9.0cm]{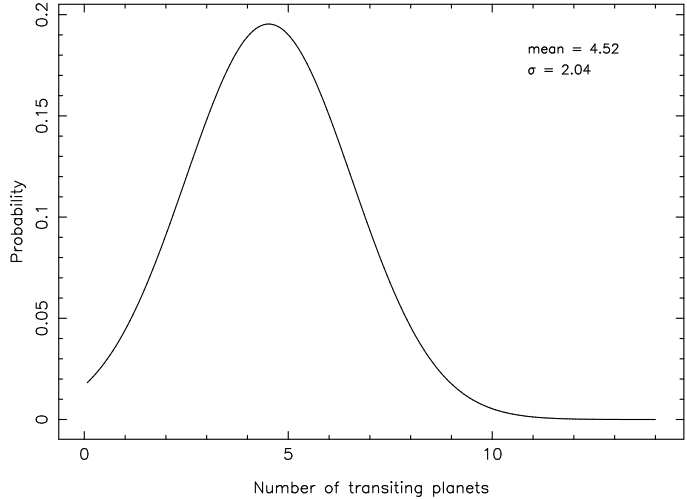}
  \end{center}
  \caption{The probability distribution for the 203 planets in the
    \cite{but06} sample, predicting the number of transiting planets
    based on their estimated orbital parameters.}
\end{figure}



\end{document}